\documentclass[aps,prb,twocolumn,superscriptaddress,floatfix,longbibliography]{revtex4-2}
\usepackage{xcolor}
\usepackage{orcidlink}
\usepackage{amsmath,amssymb} % math symbols
\usepackage{physics}
\usepackage{bm} % bold math font
\usepackage [autostyle, english = american]{csquotes}
\MakeOuterQuote{"}
\usepackage{braket}
\usepackage{enumitem}
\usepackage{graphicx} % for figures
\usepackage{comment} % allows block comments
\usepackage{textcomp} % This package is just to give the text quote '
\usepackage{enumitem}
\setlist{noitemsep,leftmargin=*,topsep=0pt,parsep=0pt}

\usepackage[normalem]{ulem}
\usepackage{xcolor} % \textcolor{red}{text} will be red for notes
\definecolor{lightgray}{gray}{0.6}
\definecolor{medgray}{gray}{0.4}

\usepackage{braket}

\usepackage{hyperref}
\hypersetup{
colorlinks=true,
urlcolor= blue,
citecolor=blue,
linkcolor= blue,
}
\newcommand{\mytitle}{Localization and Transport in a Non-Hermitian Hexagonal Harper Model}

\begin{document}

\title{\mytitle}
\author{Akshey Rajoriya\,\orcidlink{0009-0009-0999-1203}}
\email[]{akshey23@iiserb.ac.in}
\affiliation{Department of Physics, Indian Institute of Science Education and Research, Bhopal, India}

\author{Dibyajyoti Sahu\,\orcidlink{0000-0003-4739-7538}}
\email[]{dibyajyoti20@iiserb.ac.in}
\affiliation{Department of Physics, Indian Institute of Science Education and Research, Bhopal, India}

\author{Suhas Gangadharaiah\,\orcidlink{0000-0001-7834-9438}}
\email[]{suhasg@iiserb.ac.in}
\affiliation{Department of Physics, Indian Institute of Science Education and Research, Bhopal, India}
\author{Tanay Nag\,\orcidlink{0000-0001-6052-7232}}
\email[]{tanay.nag@hyderabad.bits-pilani.ac.in}
\affiliation{Department of Physics, BITS Pilani-Hyderabad Campus, Telangana 500078, India}

\date{\today}

\begin{abstract}
We investigate a one-dimensional non-Hermitian hexagonal Harper model with quasiperiodically modulated hopping amplitudes. In the Hermitian limit, the model exhibits metallic, insulating, and multifractal phases characterized by distinct eigenstate properties. Upon introducing non-Hermiticity, the phase diagram is qualitatively altered, with an expansion of metallic regions and strong boundary sensitivity arising from the non-Hermitian skin effect. By analyzing wave-packet dynamics, we uncover qualitatively distinct
transport signatures in metallic, multifractal, and insulating regimes. In the metallic region, nonreciprocal hopping induces finite sliding, resulting in ballistic center-of-mass motion that is absent in the Hermitian model,  while wave-packet spreading is simultaneously suppressed and exhibits diffusive scaling. Interestingly, the multifractal regime emerges as a distinct dynamical phase supporting both enhanced spreading and finite sliding, both primarily of superdiffusive nature, in contrast to metallic regions where sliding (spreading) shows ballistic (diffusive) scaling.  These features are markedly different from their Hermitian counterpart.  On the other hand, in the insulating region, both the spreading and sliding are strongly suppressed. We reconfirm these intriguing transport characteristics by investigating the distinct growth profile of single-particle entanglement entropy where the effect of spreading of the wave-packet is clearly manifested.
These results demonstrate that quasiperiodicity in hopping amplitudes, combined with non-Hermiticity, establishes the multifractal regime as a key mediator of transport.
\end{abstract}

\maketitle

\section{Introduction}

The interplay between localization and disorder has remained a central topic of research in condensed matter physics and quantum transport since Anderson’s seminal discovery that disorder can suppress wave propagation through interference effects even in the absence of any many-body interaction \cite{anderson1958absence}. The subsequent scaling theory of localization established that in one and two dimensions, an arbitrarily weak random disorder localizes all single-particle states in the thermodynamic limit \cite{abrahams1979scaling,evers2008anderson}. Beyond random disorder, quasiperiodic systems provide a deterministic route to localization through incommensurate modulations in the potential. In particular, the Aubry-André model exhibits a localization transition at a finite modulation strength separating extended and localized phases even in one dimension \cite{aubry1980analyticity}. Subsequent generalizations revealed richer phenomena arising from the coexistence of localized and extended states, including mobility edges, multifractal eigenstates, and anomalous transport in quasiperiodic lattices \cite{biddle2010predicted,ganeshan2015nearest,sutradhar2019transport}.

In recent years, these paradigms have been extended to non-Hermitian systems, where gain, loss, and nonreciprocal couplings fundamentally alter both spectral and dynamical properties \cite{ashida2020non,bergholtz2021exceptional,kawabata2019symmetry,sgdb-x7vx,PhysRevB.102.024205,PhysRevB.104.174501,PhysRevB.110.L041102,PhysRevB.106.214207}. Non-Hermitian systems exhibit phenomena that are absent in their Hermitian counterparts. For example, in topological systems, non-Hermiticity gives rise to the non-Hermitian skin effect and the emergence of non-Bloch band theory, thereby modifying the conventional bulk-boundary correspondence \cite{PhysRevB.106.L140303,hj3p-d7vl,PhysRevB.110.205429,PhysRevB.110.125424,PhysRevB.110.125427,PhysRevB.110.115403,PhysRevLett.123.066405,PhysRevX.8.031079}. Interestingly, non-Hermiticity induces anomalous features in dynamical quantum phase transitions that are absent in their Hermitian counterparts \cite{PhysRevB.107.184311,Mondal2024,PhysRevB.106.054308}. The non-Hermitian effects are also manifested in the localization phenomena. Unlike conventional Hermitian systems, where spectral localization and dynamical localization are intimately linked such that exponentially localized eigenstates suppress long-range transport, non-Hermitian systems can break this correspondence, as is evident from various observations \cite{PhysRevResearch.2.012074,PhysRevB.105.165114,PhysRevB.107.L220201,PhysRevE.109.044315,sgdb-x7vx,longhi2023anderson}. In particular, several studies have demonstrated that dissipative disorder or asymmetric hopping can generate situations in which localized eigenstates coexist with dynamically spreading wave packets \cite{weidemann2021coexistence,tzortzakakis2021transport,leventis2022non,longhi2023anderson,chakrabarty2023skin}. This unconventional behavior originates from the nonunitary nature of the dynamics and represents a qualitative departure from the standard Hermitian picture of localization and transport.

Beyond the breakdown of the localization–transport correspondence, recent theoretical studies have revealed that non-Hermitian systems can exhibit entirely new classes of dynamical scaling behavior. In disordered non-Hermitian lattices, wave-packet spreading may display sub-diffusive, diffusive, or sub-ballistic transport governed by universal scaling laws with no direct Hermitian analogue \cite{li2025universal,xing2025universal}. Remarkably, the transport exponents are controlled not only by eigenstate localization properties but also by statistical features of the complex spectrum, particularly the distribution of the imaginary parts of eigenenergies \cite{li2025universal}. These results indicate that transport in non-Hermitian systems is governed by the interplay between spectral structure, amplification or decay dynamics, and nonorthogonality of eigenstates.

  Parallel to these developments, quasiperiodic systems have also been generalized to non-Hermitian settings \cite{jazaeri2001localization,jiang2019interplay, longhi2019topological,liu2021exact, liu2021localization,liu2020non,schiffer2021anderson,cai2021localization,liu2021exactnh,zeng2020winding,longhi2019metal,yuce2022coexistence,cai2021boundary,xu2021non,Han2022Dimerization,Liu2022RealComplex,Chen2022Breakdown, Zhou2023NonAbelian,Xia2022Exact, Acharya2022Localization,Gandhi2023Triple, Padhan2024Complete,Lin2022Topological,Tang2021Localization,Cai2022Equivalence,Acharya2024LocalizationTransitions}. In non-Hermitian extensions of the Aubry--André--Harper model, asymmetric hopping or complex potentials lead to modified localization transitions and unconventional transport properties. Notably, the above localization-delocalization transition can become discontinuous in both the diffusion exponent and propagation velocity, and disorder can even enhance transport in certain regimes \cite{longhi2021phase}. These studies demonstrate that non-Hermiticity substantially modifies the relationship between disorder, localization, and transport, even in deterministic quasiperiodic systems. Despite these advances, most existing works have focused on disorder or quasiperiodicity introduced through onsite potentials. In contrast, the role of quasiperiodicity in the hopping amplitudes, especially in non-Hermitian systems, remains largely unexplored. This distinction is crucial, as hopping modulations directly control interference pathways and transport channels and may therefore lead to qualitatively different dynamical behavior and scaling properties compared to onsite disorder. In particular, it remains an open question how non-Hermiticity modifies localization and transport in quasiperiodic systems with hopping modulation rather than onsite modulation.

To address this question, we consider a reduced one-dimensional hexagonal Harper (HH) model with modulated hopping amplitudes \cite{Jose_2018}. We find that nonreciprocal hopping enlarges the metallic phase, while the metal-insulator transition is accompanied by a multifractal phase. By combining spectral characterization with wave-packet evolution, we show that non-Hermiticity induces directional sliding, characterized by a finite center-of-mass velocity, in contrast to the purely dispersive wave-packet spreading observed in the Hermitian system. While metallic phases exhibit strong sliding accompanied by suppressed spreading, the multifractal regime supports both directional sliding and enhanced spreading due to the sparse but spatially extended nature of multifractal eigenstates. In the insulating phase, the sliding and spreading are both substantially suppressed. 
We further demonstrate that the resulting dynamics exhibit distinct scaling behaviors across the metallic and multifractal regions: the metallic region supports ballistic center-of-mass motion alongside diffusive spreading, whereas the multifractal region exhibits primarily superdiffusive scaling in both regimes. Along these lines, we find that entanglement entropy shows different growth profiles leading to a rich phase diagram.    These results identify multifractal states as an important mediator of transport in non-Hermitian quasiperiodic systems and reveal how nonreciprocity fundamentally reshapes the relation between eigenstate structure and quantum dynamics.

The remainder of this paper is organized as follows. In Sec.~\ref{s2}, we introduce the non-Hermitian quasiperiodic hopping model, based on a one-dimensional reduction of the HH lattice, and describe the observables to characterize localization and eigenstate structure. In Sec.~\ref{s3}, we analyze the spectral properties of the system, including the phase diagram in terms of the fractal dimension, and examine the effects of non-Hermiticity under periodic boundary conditions (PBC) and open boundary conditions (OBC). In Sec.~\ref{s4}, we investigate the wave-packet dynamics, focusing on transport properties, sliding, and spreading, and characterize the associated scaling behavior across metallic, multifractal, and insulating regimes. In Sec.~\ref{s5}, we examine single-particle entanglement entropy as an independent dynamical probe of the phases. Finally, in Sec.~\ref{s6}, we summarize our results and discuss their implications for non-Hermitian quasiperiodic systems.

%%%%%%%%%%%%%%%%%%%%%%%%%%%%%%%%%%%%%%%%%%%%%%%%%%%%%%%%%%%%%%%%%%

\section{Model and Observables}
\label{s2}
%%%%%%%%%%%%%%%%%%%%%%%%%%%%%%%%%%%%%%%%%%%%%%%%%%%%%%%%%%%%%%%%%%

%%%%%%%%%%%%%%%%%%%%%%%%%%%%%%%%%%%%%%%%%%%%%%%%%%%%%%%%%%%%%%%%%%
\begin{figure}
    \centering
    \includegraphics[width=1\linewidth]{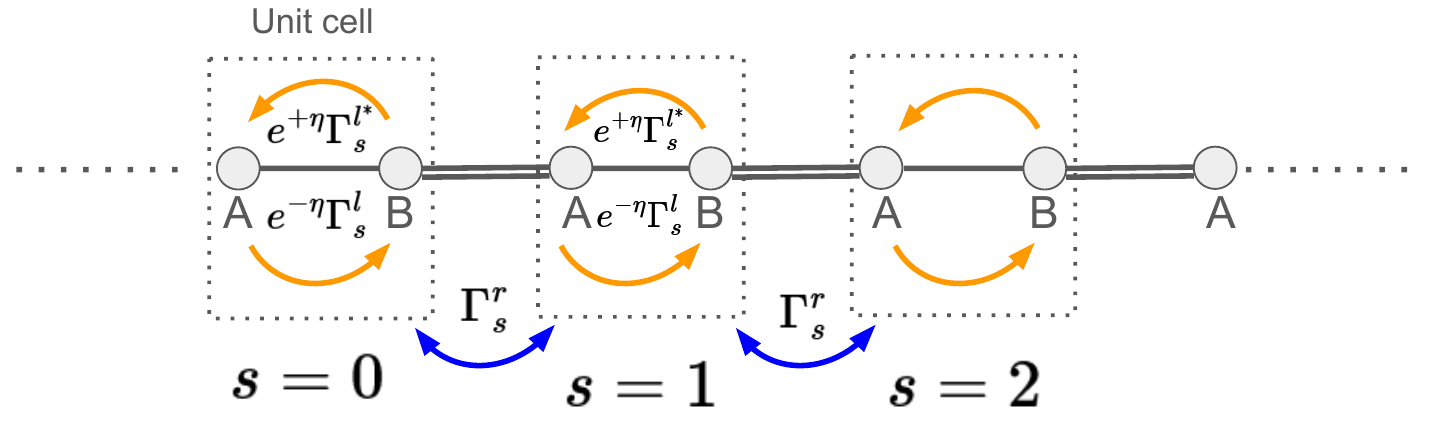}
  \caption{ Schematic of the non-Hermitian Hexagonal Harper model. Orange arrows denote the nonreciprocal intracell hopping, while blue arrows denote the reciprocal intercell hopping. }
    \label{fig:schematic}
\end{figure}

%%%%%%%%%%%%%%%%%%%%%%%%%%%%%%%%%%%%%%%%%%%%%%%%%%%%%%%%%%%%%%%%%%

We consider a non-Hermitian extension of a reduced HH model in one dimension, obtained from an anisotropic hexagonal lattice in the presence of a magnetic flux~\cite{Jose_2018,Dhara_2024}. Unlike the generalized Aubry-André-Harper (GAAH) model, in which quasiperiodicity enters through the onsite potential, the HH model incorporates quasiperiodicity exclusively via the hopping amplitudes which have a dimerized pattern. The Hamiltonian of the non-Hermitian reduced HH model is given by
\begin{eqnarray}
H &=& \sum_{s} \bigg( 
e^{\eta}\Gamma_{s}^{l} \, c^{\dagger}_{2s-1} c_{2s}
+ \Gamma_{s}^{r} \, c^{\dagger}_{2s+1} c_{2s}  \nonumber \\
 &+& e^{-\eta} \overline{\Gamma_{s}^{l}} \, c^{\dagger}_{2s} c_{2s-1}
+ \overline{\Gamma_{s}^{r}} \, c^{\dagger}_{2s} c_{2s+1}\bigg),
\label{eq:hamHH}
\end{eqnarray}
where $c^{\dagger}_{i}$ ($c_i$) denotes the creation (annihilation) operator at site $i$. The left hopping amplitude is quasiperiodic,
\begin{equation}
\Gamma_{s}^{l} = -t_1 - t_{2} e^{-i4\pi s \phi},
\end{equation}
while the right hopping amplitude is taken to be uniform, $\Gamma_s^r = -t_3$. $\overline{\Gamma}$ represents complex conjugate of $\Gamma$. Throughout this work, we fix $t_3 = 1$ and choose the incommensurate flux $\phi = (\sqrt{5}-1)/2$, unless stated otherwise. The parameter $\eta$ controls the strength of non-Hermiticity and introduces an asymmetry between forward and backward hopping processes. The schematic of the model is shown in Fig. \ref{fig:schematic}. In the Hermitian limit $\eta = 0$, the model reduces to the standard HH chain, whereas finite $\eta$ drives the system into a regime exhibiting non-Hermitian skin effects and modified localization behavior.

To characterize the eigenstate properties of the system, we analyze both PBC and OBC, which exhibit qualitatively different features in non-Hermitian systems \cite{brody2014biorthogonal}. Under PBC, the Hamiltonian admits a complete biorthogonal set of left and right eigenvectors. In this case, observables are computed within the biorthogonal framework, ensuring a consistent normalization and a proper characterization of extended and localized states. In contrast, under OBC, the system generically exhibits a strong non-Hermitian skin effect, where a macroscopic fraction of eigenstates accumulate near the boundaries \cite{yao2018edge}. In this regime, the biorthogonal basis becomes ill-conditioned and fails to capture the physically relevant spatial structure of the eigenstates. Therefore, for OBC one should restrict to either the right or left  eigenvectors. In this work we compute all observables using right eigenvectors only, which correctly reflects the spatial localization induced by the skin effect.
%%%%%%%%%%%%%%%%%%%%%%%%%%%%%%%%%%%%%%%%%%%%%%%%%%%%%%%%%%%%%%%%%%%

The localization properties are quantified using the inverse participation ratio (IPR) \cite{dominguez2019aubry,PhysRevB.83.184206,PhysRevLett.84.3690}, defined for a normalized $n$-th right eigenstate $\ket{\psi_n}$, associated with the HH model Eq. (\ref{eq:hamHH}), as
\begin{equation}
\mathrm{IPR}_n = \sum_{i=1}^{L} |\psi_n(i)|^4.
\end{equation}
For extended states, $\mathrm{IPR}_n$ becomes vanishingly small $\sim L^{-1}$, whereas for localized states it remains finite $\sim L^{0}$, in the thermodynamic limit. To further characterize the nature of eigenstates, we compute the fractal dimension $D_2$, defined through the scaling relation \cite{PhysRevB.62.7920,evers2008anderson}
\begin{equation}
\mathrm{IPR}_n \sim L^{-D_2}.
\end{equation}
Here, $D_2 = 1$ corresponds to fully extended states, $D_2 = 0$ to localized states, and intermediate values indicate multifractal or critical states. 
In addition, we compute the normalized participation ratio (NPR), defined as \cite{Kravtsov_2015}
\begin{equation}
\mathrm{NPR}_n = \left( L \sum_i |\psi_n(i)|^4 \right)^{-1},
\end{equation}
which provides a complementary measure of delocalization. For extended states, $\mathrm{NPR}_n \sim \mathcal{O}(1)$, while it approaches zero in the localized regime.

$\mathrm{NPR}$  and $D_2$ are obtained via exact diagonalization of finite systems of size $L$, and unless otherwise specified, are averaged over the full spectrum. The combined analysis using NPR and $D_2$, together with the distinction between PBC and OBC, enables a systematic investigation of the interplay between quasiperiodicity and non-Hermiticity, and allows us to identify extended, localized, and critical regimes, as well as their sensitivity to boundary-induced skin effects.

%%%%%%%%%%%%%%%%%%%%%%%%%%%%%%%%%%%%%%%%%%%%%%%%%%%%%%%%%%%%%%%%%%

\section{Localization and spectral   properties}
\label{s3}
%%%%%%%%%%%%%%%%%%%%%%%%%%%%%%%%%%%%%%%%%%%%%%%%%%%%%%%%%%%%%%%%%%

%%%%%%%%%%%%%%%%%%%%%%%%%%%%%%%%%%%%%%%%%%%%%%%%%%%%%%%%%%%%%%%%

\begin{figure}[t]
    \centering
    \includegraphics[width=1\linewidth]{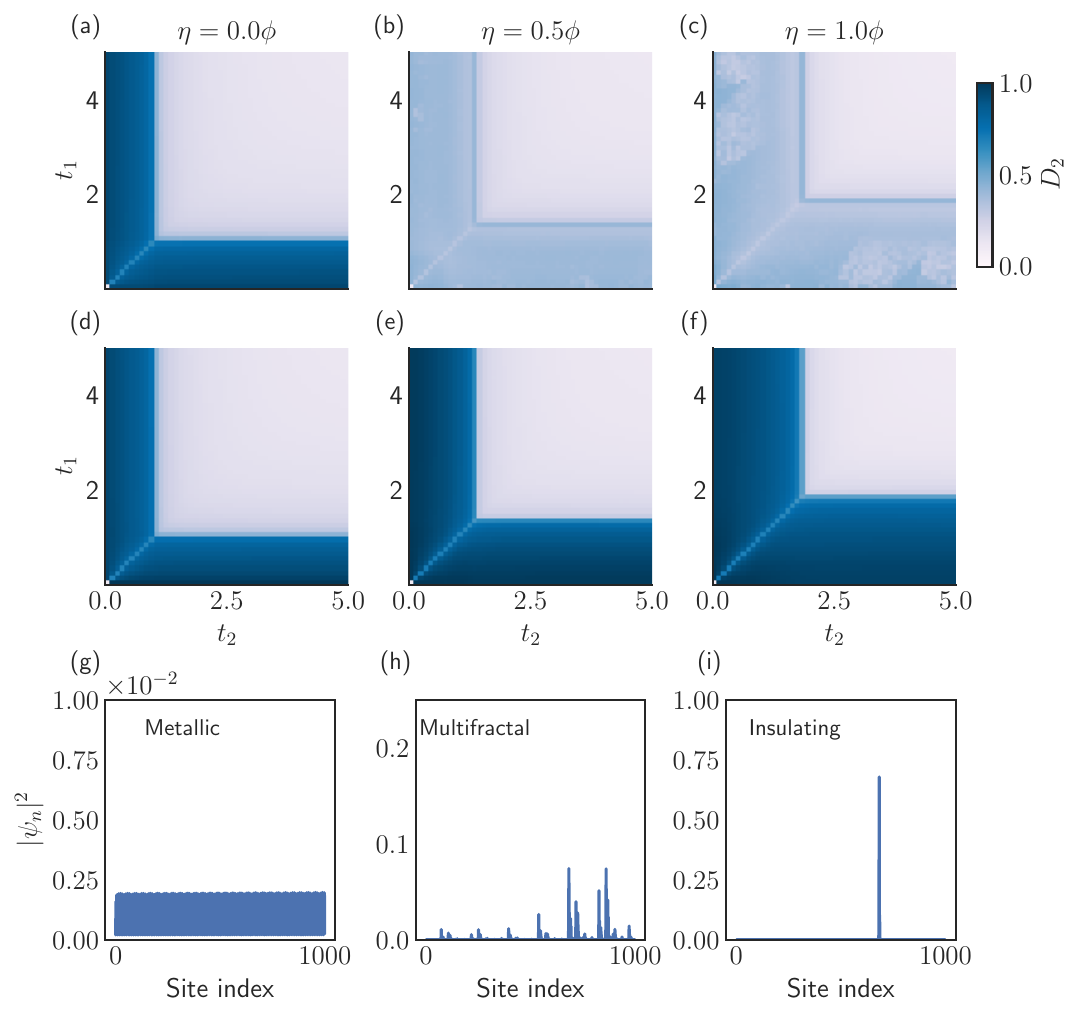}
    \caption{Average fractal dimension $D_2$ in the $(t_1,t_2)$ plane for different non-Hermitian strengths $\eta$ (in units of the golden ratio $\phi$) with $t_3=1$. Panels (a)--(c) and (d)--(f) show the phase diagrams under OBC and PBC, respectively, for $\eta=0$, $0.5\phi$, and $\phi$. Panels (g)--(i) show representative eigenstate probability distributions $|\psi_n|^2$ under PBC at $(t_1,t_2)=(3.0,0.1)$, $(3.0,1.81)$, and $(3.0,4.5)$, corresponding to the metallic, multifractal, and insulating phases, respectively. The system size is $L=1000$. 
    }
    \label{fig:nonherm_phase}
\end{figure}

%%%%%%%%%%%%%%%%%%%%%%%%%%%%%%%%%%%%%%%%%%%%%%%%%%%%%%%%%%%%%%%%

\begin{figure*}[t]
    \centering

    \includegraphics[width=1\linewidth]{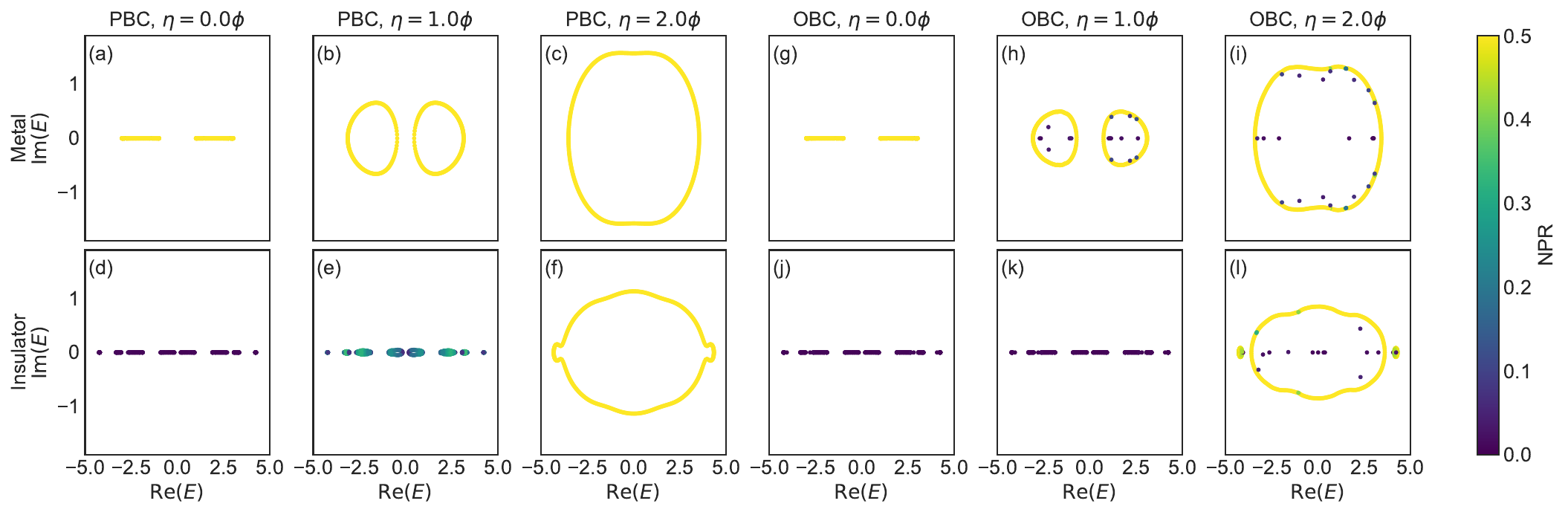}

    \caption{ Complex energy spectrum in the $\mathrm{Re}(E)$--$\mathrm{Im}(E)$ plane, with eigenstates colored by the NPR. Panels (a)--(f) [(g)--(l)] correspond to PBC (OBC) for $\eta=0,\phi,2\phi$. Panels (a)--(c) and (g)--(i) show a metallic point $(t_1,t_2)=(2.0,0.1)$, while panels (d)--(f) and (j)--(l) show an insulating point $(t_1,t_2)=(2.0,1.81)$, with $t_3=1$. In the metallic regime, the spectrum evolves from a purely real structure to complex point-gap structures, while the pronounced differences between PBC and OBC spectra indicate the emergence of the NHSE. In the insulating regime, increasing non-Hermiticity drives a transition toward extended states accompanied by boundary accumulation under OBC. The system size is $L=1000$.}
    \label{fig:spectrum}
\end{figure*}

%%%%%%%%%%%%%%%%%%%%%%%%%%%%%%%%%%%%%%%%%%%%%%%%%%%%%%%%%%%%%%%%
We characterize the localization properties of the system by computing the average fractal dimension $D_2$ from the finite-size scaling of the inverse participation ratio (IPR). Figures~\ref{fig:nonherm_phase}(a--c) and (d--f) show the resulting $D_2$ phase diagrams in the $(t_1,t_2)$ plane for different values of the non-Hermitian parameter $\eta$ under OBC and PBC, respectively. In the Hermitian limit ($\eta=0$), the phase diagram consists of two metallic phases ($D_2\approx1$) and an insulating phase ($D_2\approx0$), separated by critical lines supporting multifractal states that intersect at the bicritical point $(t_1,t_2)=(1,1)$. As the non-Hermitian parameter is increased from $\eta=0$ to $\phi$, the critical lines shift, resulting in a systematic expansion of the metallic region under both boundary conditions.

Although the overall evolution of the phase boundaries is similar under OBC and PBC, the localization properties of the eigenstates differ significantly. Under OBC [Figs.~\ref{fig:nonherm_phase}(a--c)], increasing non-Hermiticity induces the non-Hermitian skin effect (NHSE), causing eigenstates to accumulate near the system boundaries. Consequently, the average fractal dimension remains below unity even within the nominally metallic region, and the metallic phase develops a fragmented appearance that reflects the highly nonuniform spatial structure of the skin modes. In contrast, under PBC [Figs.~\ref{fig:nonherm_phase}(d--f)], where the NHSE is absent, the metallic region maintains a nearly uniform value of $D_2\approx1$, indicating genuinely extended eigenstates despite the continuous shift of the phase boundaries with increasing non-Hermiticity.

Representative eigenstate probability distributions corresponding to the metallic, multifractal, and insulating phases are shown in Figs.~\ref{fig:nonherm_phase}(g--i). The metallic state [Fig.~\ref{fig:nonherm_phase}(g)] exhibits an almost uniform probability distribution over the entire lattice, characteristic of an extended state. The multifractal state [Fig.~\ref{fig:nonherm_phase}(h)] displays pronounced amplitude fluctuations over multiple length scales, reflecting its critical nature. In contrast, the insulating state [Fig.~\ref{fig:nonherm_phase}(i)] is strongly localized, with the probability concentrated within a small spatial region. These representative wave function profiles are fully consistent with the phase classification inferred from the $D_2$ phase diagrams.

To further explore the spectral structure, we examine the complex energy spectrum by plotting $\mathrm{Re}(E)$ versus $\mathrm{Im}(E)$, with the eigenstates colored according to the NPR, as shown in Fig.~\ref{fig:spectrum}. We select representative points from the metallic and insulating regions of the Hermitian phase diagram and track their evolution as the non-Hermiticity is increased from $\eta = 0$ to $2\phi$ at an interval of $\phi$. Figures~\ref{fig:spectrum}(a--f) and \ref{fig:spectrum}(g--l) show the results under PBC and OBC, respectively, with panels (a--c) and (g--i) corresponding to the metallic phase, and panels (d--f) and (j--l) corresponding to the insulating phase. For parameters residing within the metallic phase, the spectrum under PBC evolves from a purely real line to structures exhibiting both line-gap and point-gap features at intermediate non-Hermiticity ($\eta = \phi$), eventually forming a single point-gap at larger $\eta$. Under OBC, the spectral structure remains largely unchanged; however, a subset of states that are extended under PBC develop low-NPR signatures due to NHSE, indicating boundary localization even within regions that remain metallic in PBC. For parameters taken within the insulating phase, the spectrum remains essentially unchanged at moderate non-Hermiticity ($\eta = \phi$), retaining its localized character; however, at stronger non-Hermiticity ($\eta = 2\phi$), the states become extended, with a fraction exhibiting NHSE under OBC. This demonstrates that the insulating phase of the Hermitian model is destabilized by non-Hermiticity and evolves into a metallic regime, consistent with the expansion of metallic regions observed in the $D_2$ phase diagram.
%%%%%%%%%%%%%%%%%%%%%%%%%%%%%%%%%%%%%%%%%%%%%%%%%%%%%%%%%%%%%%%%%%

\section{Wave-packet dynamics}
\label{s4}
%%%%%%%%%%%%%%%%%%%%%%%%%%%%%%%%%%%%%%%%%%%%%%%%%%%%%%%%%%%%%%%%%%

%%%%%%%%%%%%%%%%%%%%%%%%%%%%%%%%%%%%%%%%%%%%%
We next analyze the wave-packet dynamics to characterize the transport properties of the system \cite{longhi2021phase,orito2022unusual}. We consider an initial state localized at a single lattice site, $\ket{\psi(t=0)} = \ket{j_0}$, and evolve it under the Hamiltonian as
\begin{equation}
\ket{\psi(t)} = e^{-iHt}\ket{\psi(0)}.
\end{equation}
In non-Hermitian systems, the norm of the wave function is generally not conserved due to the complex energy spectrum. To maintain a consistent probabilistic interpretation, we normalize the time-evolved state at each instant as
\begin{equation}
\ket{\tilde{\psi}(t)}=
\frac{\ket{\psi(t)}}{\sqrt{\langle\psi(t)|\psi(t)\rangle}},
\end{equation}
where $\ket{\psi(t)}$ denotes the time-evolved right state. Under OBC, $\bra{\psi(t)}=\ket{\psi(t)}^{\dagger}$, whereas under PBC, $\bra{\psi(t)}$ denotes the corresponding time-evolved left state, so that the normalization and the probability density are evaluated using the biorthogonal inner product. The dynamics is then characterized by the spatial probability distribution $|\tilde{\psi}_j(t)|^2$, from which we define the center of mass
\begin{equation}
x_{CM}(t) = \sum_j j\, |\tilde{\psi}_j(t)|^2,
\end{equation}
with the associated sliding velocity, $v_s = [x_{CM}(t) - x_{CM}(0)]/t$, capturing directed transport. The spread of the wave packet is defined as
\begin{equation}
\Delta x(t) = \sqrt{\sum_j \left[j - x_{CM}(t)\right]^2 |\tilde{\psi}_j(t)|^2},
\end{equation}
and its growth is quantified by
\begin{equation}
\sigma(t) = \Delta x(t) - \Delta x(0).
\end{equation}
Together, $v_s$ and $\sigma(t)$  \cite{orito2022unusual} provide complementary measures of sliding and spreading, respectively. Note that sliding velocity effectively measures the phase velocity while wave-packet spreading is connected with the group velocity.

\begin{figure}[t]
    \centering
    \includegraphics[width=1\linewidth]{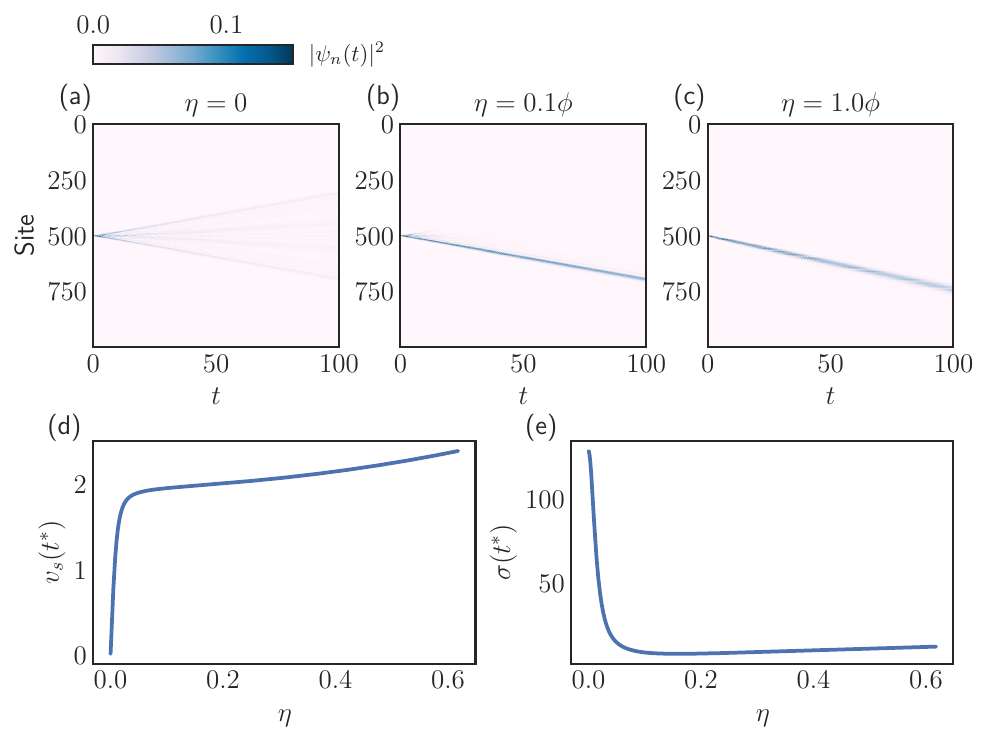}
\caption{Spatiotemporal evolution of a wave packet in the non-Hermitian HH model for $t_1=3.0$ and $t_2=0.1$. Panels (a)--(c) show the probability density $|\psi_j(t)|^2$ for $\eta=0$, $0.1\phi$, and $1.0\phi$, respectively. Panels (d) and (e) show the sliding velocity $v_s(t^\ast)$ and wave-packet width $\sigma(t^\ast)$ as functions of $\eta$ at $t^\ast=100$. The system size is $L=1000$.}
    \label{fig:wpexmp}
\end{figure}
The real-space evolution of the wave packet in the metallic phase is shown in Figs.~\ref{fig:wpexmp}(a--c). As the system becomes increasingly non-Hermitian, the initially symmetric ballistic light cone evolves into a unidirectionally propagating wave packet due to nonreciprocal hopping, with one branch of the light cone becoming progressively suppressed. We further quantify this behavior by analyzing the sliding velocity and wave-packet spreading as functions of the nonreciprocity parameter $\eta$ in Figs.~\ref{fig:wpexmp}(d,e), respectively. The wave packet consequently acquires a finite center-of-mass velocity, while its spatial spreading is strongly suppressed. Correspondingly, the sliding velocity increases rapidly with $\eta$ before exhibiting a gradual increase at larger nonreciprocity, whereas the spreading decreases sharply upon introducing non-Hermiticity, reaches a minimum at intermediate $\eta$, and then exhibits a slight recovery for stronger nonreciprocity. In contrast, the Hermitian system exhibits symmetric ballistic expansion with zero center-of-mass velocity, resulting in the familiar symmetric light-cone profile.

\begin{figure}[t]
    \centering
    \includegraphics[width=1\linewidth]{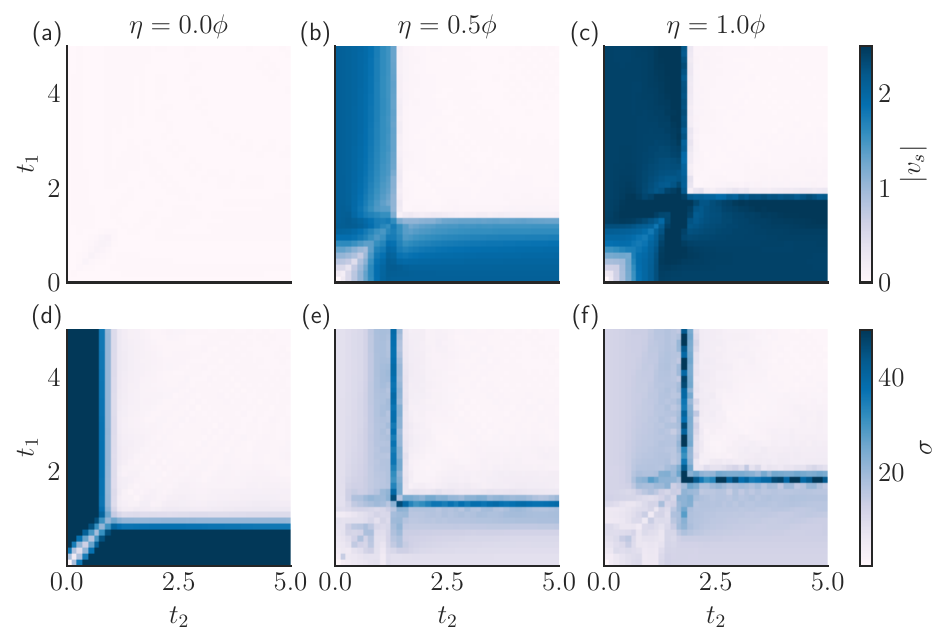}
    \caption{Wave-packet transport under OBC. Heatmaps of the sliding velocity $|v_s|$ (a--c) and wave-packet spread $\sigma$ (d--f) in the $(t_1,t_2)$ plane for $\eta=0$, $0.5\phi$, and $\phi$, respectively. Finite sliding develops with increasing non-Hermiticity, while the wave-packet spread is largest in the multifractal regime. We consider $t=100$. The system size is  $L=1000$.}
    \label{fig:OBC_trans}
\end{figure}

We study the wave-packet dynamics under OBC by plotting the heatmaps of the sliding velocity $v_s$ and the spread $\sigma$ in the $(t_1,t_2)$ plane for increasing non-Hermiticity $\eta = 0, 0.5\phi$ and $\phi$ (see Fig.~\ref{fig:OBC_trans}). In the Hermitian limit $\eta = 0$, see Figs.~\ref{fig:OBC_trans}(a,d), the dynamics is governed solely by quasiperiodic modulation, and the sliding velocity vanishes across the phase diagram, reflecting the absence of nonreciprocal transport. The spreading, however, clearly distinguishes different phases: $\sigma$ is finite in metallic regions, indicating extended dynamics, while it decreases and eventually vanishes upon approaching the insulating regime, consistent with localization. Upon introducing non-Hermiticity with $\eta = 0.5\phi$, see Figs.~\ref{fig:OBC_trans}(b,e), a qualitative change in the dynamics is observed. A finite sliding velocity develops in the metallic regions, signaling directional transport induced by nonreciprocal hopping. At the same time, the spreading is strongly suppressed compared to the Hermitian case, even within regions that were previously extended. In contrast, the largest spreading now occurs in the multifractal regions, sandwiched between metallic and insulating regions,
where both $v_s$ and $\sigma$ are appreciable. This indicates a coexistence of sliding and spreading in the multifractal region. For stronger non-Hermiticity with $\eta = \phi$, see Figs.~\ref{fig:OBC_trans}(c,f), the above features persist with only minor quantitative changes. 
% The metallic regions continue to exhibit finite sliding velocity with reduced spreading, while the multifractal regions remain the dominant contributors to wave-packet broadening. 
This shows that non-Hermiticity reorganizes the dynamics, with sliding and spreading playing distinct roles: metallic regions are characterized by finite sliding with suppressed spreading, whereas the multifractal regime supports both significant sliding and enhanced spreading. Since the dynamics is tracked only until the wave packet reaches the system boundary, the same qualitative behavior is observed under PBC.

\begin{figure}
    \centering
    \includegraphics[width=1\linewidth]{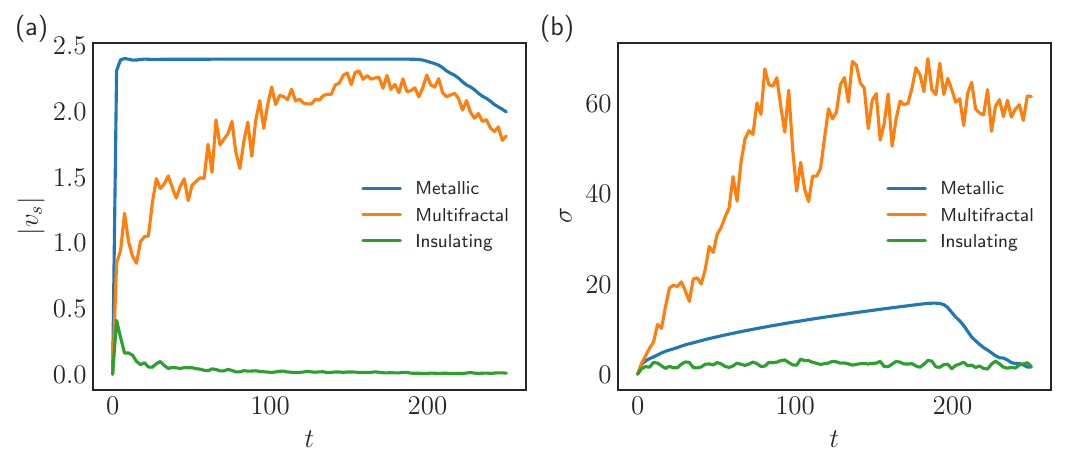}
\caption{Wave-packet transport dynamics in the non-Hermitian HH model under OBC. Time evolution of (a) the sliding velocity $|v_s|$ and (b) the wave-packet width $\sigma$ for three representative parameter sets: $(t_1,t_2)=(3.0,0.1)$, $(3.0,1.81)$, and $(3.0,4.5)$, corresponding to the metallic, multifractal, and insulating phases, respectively. The system size is $L=1000$.}
    \label{fig:vt_sigma}
\end{figure}

\begin{figure}
    \centering
    \includegraphics[width=1\linewidth]{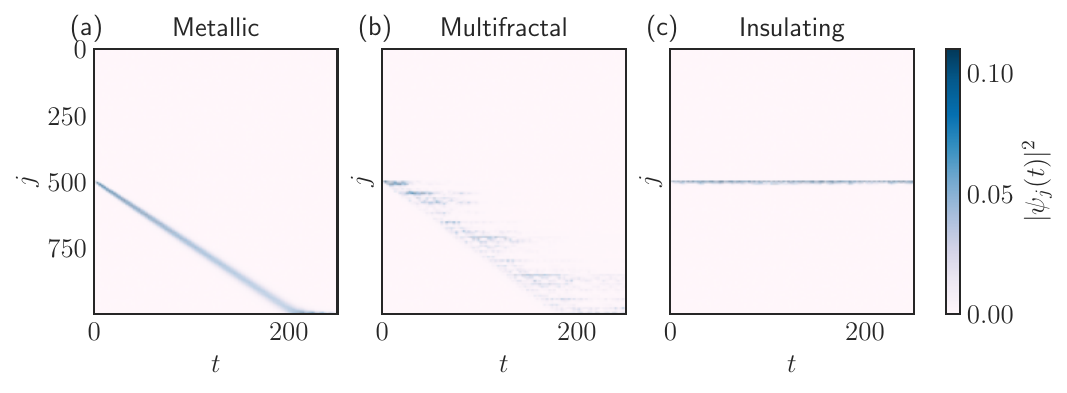}
    \caption{Spatiotemporal evolution of the wave-packet density $|\psi_j(t)|^2$ under OBC. (a) Metallic, $(t_1,t_2)=(3.0,0.1)$; (b) multifractal, $(t_1,t_2)=(3.0,1.81)$; and (c) insulating, $(t_1,t_2)=(3.0,4.5)$ phases. The system size is $L=1000$.}  
    \label{fig:wp_compare}
\end{figure}

We further examine the time evolution of $v_s$ and $\sigma$ for representative points across different regions of the phase diagram, as shown in Figs.~\ref{fig:vt_sigma}(a,b), respectively. These results clearly highlight the distinction between dynamical regimes, in particular the enhanced spreading in the multifractal region. This is further complemented by direct visualization of the wave-packet evolution in the  metallic,  multifractal, and  insulating phases, as shown in Figs.~\ref{fig:wp_compare}(a--c), respectively. Together, these observations emphasize the special role of the multifractal regime, where the eigenstates are neither fully extended nor localized but instead occupy a sparse, spatially distributed over multiple sites, as reflected in intermediate values of $D_2$. In the presence of nonreciprocal hopping, this structure enables the wave packet to propagate through spatially separated high-weight regions, leading to enhanced spreading compared to both metallic and insulating phases. 

\begin{figure}
    \centering
    \includegraphics[width=1\linewidth]{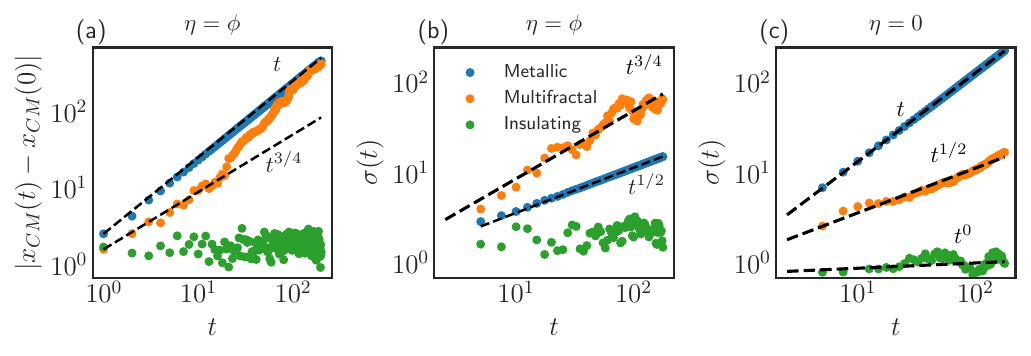}
    \caption{Dynamic wave-packet signatures across the metallic $(t_1,t_2)=(3.0,0.1)$, multifractal $(3.0,1.81)$, and insulating $(3.0,4.5)$ phases under OBC. (a) Time evolution of the center-of-mass displacement $|x_{\mathrm{CM}}(t)-x_{\mathrm{CM}}(0)|$. (b) Time evolution of the wave-packet width $\sigma(t)$ for the non-Hermitian case ($\eta=\phi$). (c) Time evolution of the wave-packet width $\sigma(t)$ for the Hermitian case ($\eta=0)$, where the multifractal phase corresponds to $(t_1,t_2)=(3.0,1.0)$. Dashed lines indicate the corresponding power-law fits. The system size is $L=1000$.}
    \label{fig:scaling}
\end{figure}

We finally turn to the scaling behavior of the center-of-mass displacement and the wave-packet spread, shown in Fig.~\ref{fig:scaling}. All scaling exponents are extracted from the pre-asymptotic time window to isolate the bulk dynamics before the onset of the non-Hermitian skin effect or boundary reflections.

The time evolution of the center-of-mass displacement $|x_{CM}(t)-x_{CM}(0)|$, shown in Fig.~\ref{fig:scaling}(a), distinguishes the three phases. The dashed and dotted black lines, associated with metallic and multifractal regions, respectively,  denote the reference power-law scalings $t$ (ballistic) and $t^{3/4}$, respectively. In the insulating phase, the center of mass remains stationary ($\sim t^0$), whereas in the metallic phase it increases linearly with time ($\sim t$), reflecting ballistic transport arising from the finite sliding velocity induced by nonreciprocal hopping, a feature absent in the Hermitian model. In contrast, the multifractal phase exhibits a crossover from superdiffusive/sub-ballistic scaling ($\sim t^{3/4}$) at intermediate times to ballistic behavior ($\sim t$) at longer times. This crossover originates from the multifractal nature of the wave functions and signals the onset of directed transport.

In contrast, the time evolution of the wave-packet width $\sigma(t)$, shown in Fig.~\ref{fig:scaling}(b), exhibits diffusive scaling ($\sim t^{1/2}$) in the metallic phase. In the multifractal phase, however, $\sigma(t)$ displays pronounced oscillations about an average power-law growth. The mean profile follows a superdiffusive scaling, $\sigma(t)\sim t^{3/4}$, corresponding to a larger exponent than the diffusive metallic behavior. The enhanced spreading originates from the delocalized character of the multifractal wave functions, whereas the oscillations reflect interference among their multiple localization centers.
 
For comparison, the corresponding wave-packet dynamics in the Hermitian limit is shown in Fig.~\ref{fig:scaling}(c), where the dashed lines denote the fitted power-law exponents for the three phases. In the metallic phase, the wave-packet width grows ballistically, $\sigma(t)\sim t$, in sharp contrast to the diffusive scaling, $\sigma(t)\sim t^{1/2}$, observed in the non-Hermitian case. This suppression of wave-packet spreading under OBC is a direct consequence of the non-Hermiticity.

Despite the above scaling features, the magnitude of $\sigma(t)$ differs significantly: the metallic phase shows strongly reduced spreading, whereas the multifractal regime exhibits enhanced spreading. This distinction originates from the underlying eigenstate structure: metallic states, although extended, are effectively directed by nonreciprocal hopping, leading to dominant sliding with limited broadening, while multifractal states occupy a sparse but spatially distributed set of sites, enabling the wave packet to propagate between multiple high-weight regions and thereby producing both larger center-of-mass displacement and enhanced spreading. Thus, non-Hermiticity reorganizes the dynamics by rendering the sliding ballistic while keeping the spreading diffusive, with the multifractal regime uniquely supporting both large displacement and maximal wave-packet broadening.

\section{Single-Particle Entanglement Entropy}
\label{s5}
Having established the spectral and transport signatures of the different phases, we now investigate whether single-particle entanglement entropy provides an independent dynamical diagnostic of the underlying phases. To this end, we calculate the single-particle entanglement entropy (SPEE) for the non-Hermitian HH model under PBC. Since the wave-packet dynamics is dominated by rapid sliding, OBC can obscure the long-time spreading behavior in the strongly suppressed metallic regime. We therefore employ PBC to minimize boundary effects and capture the intrinsic long-time spreading dynamics. Unlike conventional many-body entanglement entropy, the quantity considered here is obtained from the dynamics of a single-particle wave packet evolving under the Hamiltonian \cite{shekhar2025single,Calabrese_2005}. The single-particle entanglement entropy is evaluated from the normalized time-evolved wave packet obtained using the evolution protocol described in Sec.~IV.

The lattice is divided into two equal subsystems,
$A=\{1,\ldots,L/2\}$ and
$B=\{L/2+1,\ldots,L\}$.
The probability of finding the particle in subsystem $A$ is
\begin{equation}
p_A(t)=\sum_{i\in A} |\tilde{\psi}_i(t)|^2,
\end{equation}
with $p_B(t)=1-p_A(t)$.
The reduced density matrix of the subsystem $A$ is obtained by tracing over subsystem $B$,
\begin{equation}
\rho_A(t) = \mathrm{Tr}_B \left(|\tilde{\psi}(t)\rangle
\langle\tilde{\psi}(t)|\right).
\end{equation}
The corresponding von Neumann entropy is
\begin{equation}
S(t)
=
-\mathrm{Tr}
\left(
\rho_A(t) \ln \rho_A(t)
\right),
\end{equation}
which, for a single-particle state, reduces to
\begin{equation}
S(t) = -p_A(t)\ln p_A(t) -[1-p_A(t)]\ln [1-p_A(t)].
\label{eq:SPEE}
\end{equation}
To characterize the entanglement dynamics at a given parameter point $(t_1,t_2)$, we compute the time-dependent single-particle entanglement entropy $S(t)$ for site-localized initial states. Unless otherwise stated, the time-dependent entropy shown in the line plots is averaged over all initial wave-packet positions to eliminate the dependence on the choice of the initial site under PBC. To construct the phase diagram, we further evaluate the long-time average,

\begin{equation}
\langle S \rangle =
\frac{1}{T-T_{\mathrm{o}}}
\int_{T_{\mathrm{o}}}^{T}
S(t)\,dt,
\end{equation}

where $T_{\mathrm{o}}$ is chosen sufficiently large such that only the post-transient dynamics contribute to the time average. The resulting long-time- and initial position-averaged entropy is then used to construct the phase diagram.

\begin{figure}
    \centering
    \includegraphics[width=1\linewidth]{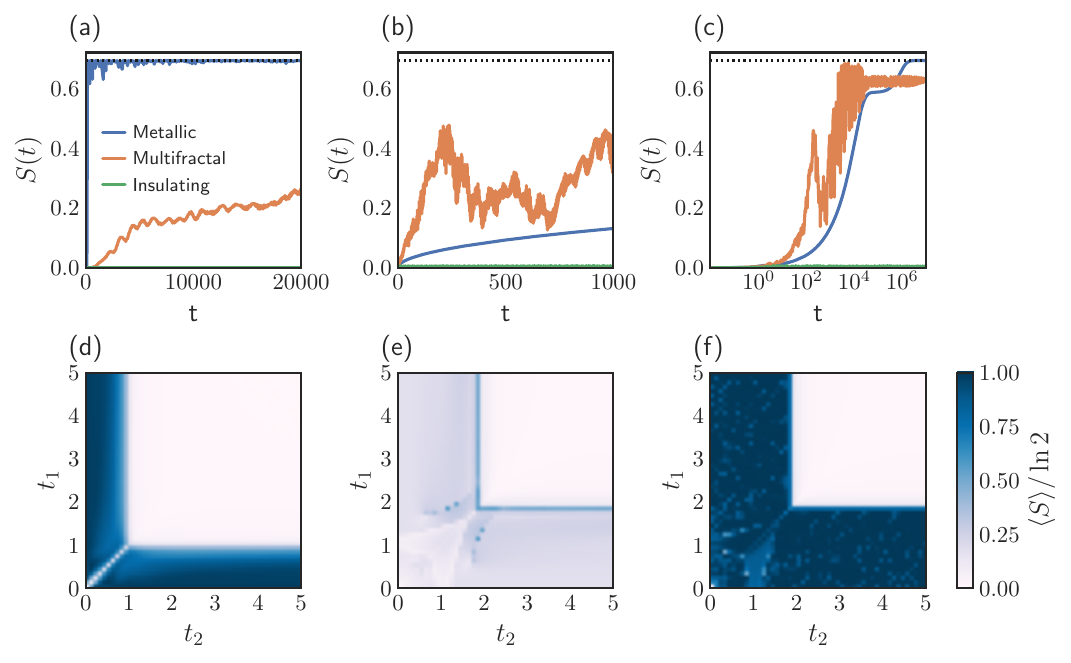}
\caption{Half-chain single-particle entanglement entropy (SPEE) for the system under OBC. (a) Hermitian dynamics ($\eta=0$), (b) early-time evolution ($t_{\max}=L$), and (c) long-time evolution of $S(t)$ for the non-Hermitian HH model with $\eta=\phi$ across representative metallic ($t_1=3.0$, $t_2=0.1$), multifractal ($t_1=3.0$, $t_2=1.81$), and insulating ($t_1=3.0$, $t_2=4.5$) phases. The horizontal dotted line indicates $\ln 2$. (d) Hermitian phase diagram at $\eta = 0$. (e) Non-Hermitian short-time phase diagram of the normalized time-averaged entropy $\langle S \rangle/\ln2$ obtained with $t_{\max}=L$. (f) Same as (e), but for the long-time limit with $t_{\max}=10^4 L$.}
    \label{fig:SPEE}
\end{figure}
Figure~\ref{fig:SPEE} summarizes the dynamical behavior of the half-chain SPEE. The upper panels [Figs.~\ref{fig:SPEE}(a)--\ref{fig:SPEE}(c)] show the initial-position-averaged entanglement entropy $S(t)$ as a function of time, while the lower panels [Figs.~\ref{fig:SPEE}(d)--\ref{fig:SPEE}(f)] present phase diagrams obtained from the long-time- and initial-position-averaged entropy $\langle S \rangle$. The Hermitian results ($\eta=0$) are shown in Figs.~\ref{fig:SPEE}(a) and \ref{fig:SPEE}(d). In the Hermitian metallic region, the SPEE $S(t)$ increases very rapidly $S(t) \propto t$ for $t<L/v_g$ with $v_g$ being the group velocity, followed by a saturation $S(t) \approx ln(2)$ for  $t>L/v_g$, which is commonly observed for ground-state entanglement entropy under unitary dynamics \cite{calabrese2005evolution,blass2012quantum,PhysRevE.97.042108,russomanno2016entanglement}. In the multifractal phase of the Hermitian model, $S(t)$ grows slowly  $ \propto t^\beta$ with $\beta<1$ while this growth is observed for a significantly large window of time as the group velocity $v_g$ decreases substantially with respect to metallic region. This observation further confirms that spreading is less in the multifractal phase as compared to the metallic region, as shown in Fig. \ref{fig:scaling} (c). On the other hand, $S(t)$ freezes in the insulating region as the particle remains localized with time.  Importantly, the SPEE acquires a finite value only when the time-evolved wave-function spreads over the two sub-subsystems. In the infinite time limit, the SPEE obeys volume and area laws with sub-system size for metallic and insulating regions, respectively \cite{hastings2007area,RevModPhys.82.277} while multifractality could lead to sub-extensive scaling \cite{PhysRevB.77.014208,PhysRevResearch.2.012074}.

For the non-Hermitian system, the early-time dynamics [Fig.~\ref{fig:SPEE}(b)] clearly distinguish the three phases. As opposed to the previous Hermitian case, the multifractal phase exhibits the fastest growth of entanglement entropy compared to that of a metallic phase. This reflects a rapid spreading of wave-packet while retaining strong spatial amplitude fluctuations. Interestingly,  the metallic phase displays comparatively slower but steady entropy growth because nonreciprocal hopping converts symmetric expansion into directional sliding, thereby reducing the effective spatial spreading. 
In the metallic phase, the SPEE growth slows down with time in the early time, $S(t)\propto t^\gamma$  with $\gamma<1$, for non-Hermitian model as compared to the Hermitian counterpart. This reduction is directly caused by less spreading with time $\sigma(t) \propto t^{1/2}$ in the non-Hermitian model as compared to the Hermitian one $\sigma(t) \propto t$, see Figs. \ref{fig:scaling} (b,c).  This is consistent with the distinct growth profile of SPEE observed for the non-Hermitian model, as reported in the literature \cite{le2023volume,j1r7-82ft}.  On the other hand, spreading is faster in the non-Hermitian multifractal phase $\sigma(t) \propto t^{3/4}$ as compared to the Hermitian multifractal one $\sigma(t) \propto t^{1/2}$, resulting in quicker growth of SPEE for the former phase in the early time limit. 
The insulating phase remains weakly entangled throughout the evolution, consistent with the localized nature of its eigenstates.

% The long-time evolution shown in Fig.~\ref{fig:SPEE}(c) reveals saturation characteristics similar to those of the Hermitian model. The distinct spreading profiles, associated with respective group velocities, across the different phases lead to different characteristic times at which the SPEE growth approaches saturation. This applies to the behavior inside the identical phase coming from  Hermitian and non-Hermitian models.
% As the dynamics proceed, the entropy in the metallic phase continues to increase and eventually approaches the values attained in the multifractal regime, indicating that the extended wave packet gradually explores the entire system. By contrast, the insulating phase maintains a consistently low entropy, reflecting the absence of significant spatial spreading. The non-Hermiticity causes pronounced oscillations in SPEE as compared to its Hermitian counterpart. 

The long-time evolution shown in Fig.~\ref{fig:SPEE}(c) reveals saturation characteristics similar to those of the Hermitian model. The distinct spreading profiles, associated with the respective group velocities across the different phases, lead to different characteristic times at which the SPEE growth approaches saturation. This is consistent with the behavior observed within the corresponding phases of the Hermitian and non-Hermitian models. In the metallic phase, the crossover toward saturation has a more subtle origin. The initially localized wave packet has finite overlap with many fixed bulk eigenstates, whose relative contributions evolve differently in time according to their imaginary energies. Consequently, although the wave packet begins to spread across the system, at intermediate times a particular bulk eigenstate, or a set of bulk eigenstates, can become dominant. Even though these states are extended, their combined probability density remains  spatially nonuniform across the entire system, leading to a metastable regime in which the SPEE saturates at a value below $\ln(2)$. At longer times, one of them or another bulk eigenstate with a larger surviving contribution can become dominant. If this eigenstate has uniform probability distribution throughout the system, the SPEE eventually escapes the metastable regime and crosses over to its maximum value, $\ln(2)$. By  contrast, in the multifractal phase, the eigenstates themselves remain multifractal. Thus, even when a particular eigenstate or a set of eigenstates dominates at long times, their probability distributions retain their multifractal character and do not extend uniformly across the entire system. Consequently, the SPEE does not reach its maximal value $\ln(2)$, although it can approach it closely, with the saturation value reaching approximately $0.6 < \ln(2)\simeq 0.693$ in our simulations. The insulating phase, on the other hand, maintains a consistently low entropy, reflecting the absence of significant spatial spreading. Thus, despite the enhanced oscillations in the SPEE introduced by non-Hermiticity compared with its Hermitian counterpart, the distinct long-time saturation behavior provides a clear means of distinguishing the metallic, multifractal, and insulating phases.

The corresponding non-Hermitian phase diagrams obtained from the normalized long-time- and initial-position-averaged entropy are presented in Figs.~\ref{fig:SPEE}(e) and \ref{fig:SPEE}(f). Using a short averaging window ($t_{\max}=L$), the multifractal region is prominently highlighted by its enhanced entropy production [Fig.~\ref{fig:SPEE}(e)], providing a clear dynamical signature of the intermediate phase. 
For a longer averaging window ($t_{\max}=10^4L$), the entropy becomes uniformly large throughout the extended region of the phase diagram [Fig.~\ref{fig:SPEE}(f)]. Some of the patches observed in the metallic regime can be attributed to the metastable behavior discussed in Fig.~\ref{fig:SPEE}(c), where the SPEE remains temporarily trapped below its maximal value. At sufficiently long times, however, the wave packet is expected to become dominated by a single extended eigenstate with a nearly uniform probability distribution, causing the entropy to approach $\ln(2)$. Thus, the nonuniform patches in the metallic regime are expected to diminish and eventually disappear at sufficiently long times making the metallic and multifractal phases less distinguishable while preserving a clear contrast with the localized phase, where the entropy remains strongly suppressed.

%%%%%%%%%%%%%%%%%%%%%%%%%%%%%%%%%%%%%%%%%%%%%%%%%%%%%%%%%%%%%%%%%%

\section{Conclusion}
\label{s6}
%%%%%%%%%%%%%%%%%%%%%%%%%%%%%%%%%%%%%%%%%%%%%%%%%%%%%%%%%%%%%%%%%%

In this work, we investigated a one-dimensional hexagonal Harper model with quasiperiodic modulation of the dimerized hopping amplitudes, and explored the interplay between quasiperiodicity and non-Hermiticity through spectral and dynamical probes. The quasiperiodic hopping gives rise to a rich phase diagram in the Hermitian limit, comprising metallic, insulating, and multifractal phases.

Upon introducing non-Hermiticity, we find that the primary effect is a redistribution of phases, with metallic regions expanding in parameter space. While PBCs retain a uniform characterization of extended states, OBCs reveal strong boundary sensitivity due to the non-Hermitian skin effect, leading to fragmentation of the phase diagram and deviations from conventional bulk behavior. The spectral analysis further shows the emergence of point-gap structures and the destabilization of the insulating phase towards the extended behavior.

These features are directly reflected in the wave-packet dynamics. In the Hermitian case, transport is governed by symmetric spreading with no directed motion, whereas non-Hermiticity induces finite sliding, leading to directional transport. At the same time, spreading is strongly suppressed in the metallic phase, indicating that nonreciprocity promotes directional sliding while limiting wave-packet broadening. In contrast, the multifractal regime emerges as a distinct dynamical phase, supporting both significant sliding and enhanced spreading. This behavior originates from the underlying eigenstate structure: multifractal states occupy a sparse but spatially distributed set of sites, which, in the presence of nonreciprocal hopping, enables the wave packet to propagate across spatially separated high-weight regions.

This picture is further supported by the time-resolved and scaling analysis of the dynamics. In the multifractal phase,  the center-of-mass displacement exhibits a crossover from superdiffusive/sub-ballistic to ballistic scaling, while the wave-packet spreading remains superdiffusive. The different scaling exponents are accompanied by an equally pronounced difference in magnitude. The metallic phase shows
diffusive spreading and ballistic sliding which is indeed unique to non-Hermitian system. Note that underlying Hermitian model exhibits ballistic and diffusive spreading in metallic and multifractal regions, respectively. Therefore, the non-Hermiticity not only alters eigenstate structure but also imprints its impact on scaling laws that control the transport. We further validate our findings using SPEE, demonstrating that it faithfully reproduces the wave-packet dynamics and accurately reconstructs the phase diagrams in both the short- and long-time regimes. Multifractal region exhibits faster growth of SPEE due to non-Hermiticity which is connected with the stronger spreading of the wave-packet. This is markedly different with respect to the Hermitian counterpart, where the metallic region shows strongest spreading as well as fastest SPEE growth.

Overall, our results demonstrate that non-Hermiticity qualitatively modifies the relationship between localization and transport in quasiperiodic hopping systems. By combining spectral analysis, wave-packet dynamics, and SPEE, we provide complementary diagnostics that consistently distinguish the metallic, multifractal, and insulating phases. Our work highlights the multifractal phase as a unique transport regime and establishes SPEE as an independent dynamical probe of localization in non-Hermitian quasiperiodic systems.

\begin{acknowledgments}
    TN  thanks the Advanced Research Grant (ARG) from Anusandhan National Research Foundation Grant No. ANRF/ARG/2025/003163/PS.
\end{acknowledgments}
 
\bibliography{ref}

\end{document}